\documentclass{article} 
\usepackage{amsmath,amssymb,amsbsy,amsfonts}
\makeatletter

          \@addtoreset{equation}{section}
       \makeatother
\allowdisplaybreaks 
\begin{document}
\title{\bf Noncommutative gauge theory \\
with arbitrary U(1) charges }%

\author{Yoshitaka {\sc Okumura}\\
{Department of Natural Science, 
Chubu University, Kasugai, {487-8501}, Japan}
}%
\date{}%
\maketitle
\begin{abstract}
It is well-known that the charge of fermion is 0 or $\pm1$ in the
U(1) gauge theory on noncommutative spacetime. 
Since 
the deviation from the standard model in particle physics has not yet observed,
and so there may be 
no room to incorporate the noncommutative U(1) gauge theory 
into the standard model because the quarks have fractional charges.
However, 
it is shown in this article   that
there is the noncommutative gauge theory with arbitrary charges
which symmetry is for example SU(3+1)$\ast$.
This enveloping gauge group consists of elements
$
\exp i\left\{\sum_{a=0}^8\,T^a\,\alpha^a(x,\theta)\ast+
Q\,\beta(x,\theta)\ast\right\}
$
with 
$
Q=\text{diag}(e,e,e,e')
$
and the restriction $\lim_{\,\theta\to0}\alpha^0(x,\theta)=0.$
This type of gauge theory is emergent from 
the spontaneous breakdown of 
 the noncommutative  SU(N)$\ast$ or SO(N)$\ast$ gauge theory 
in which 
the gauge field contains the 0 component $A_\mu^0(x,\theta)$.
$A_\mu^0(x,\theta)$ can be eliminated by gauge transformation.
Thus, 
the noncommutative gauge theory with arbitrary U(1) charges 
can not exist alone, but it must coexist with 
noncommutative nonabelian gauge theory.
This suggests that the spacetime noncommutativity requires the 
grand unified theory which spontaneously breaks down to 
the noncommutative standard model with fractionally charged quarks. 
\end{abstract}



\thispagestyle{empty}
\section{Introduction\label{S1}}
In the past several years, field theories on the noncommutative(NC) spacetime
have been extensively studied from many different aspects. The motivation 
comes from the string theory which makes obvious that end points of 
the open strings trapped on the D-brane in the presence of 
 two form B-field background turn out to be noncommutative 
\cite{SJ}
and 
then the noncommutative supersymmetric gauge theories appear as the low energy effective theory of such D-brane \cite{DH}, \cite{SW}.
\par
Hayakawa \cite{Hayakawa} indicated that the matter field must have charge 
0 or $\pm1$ in U(1) NC  gauge theory  in order to keep the gauge invariance of the theory.
Armoni \cite{Armoni} also indicated that U(N) gauge theory has the consistency in calculations of gluon propagator and  three gluons vertex to one loop order, whereas SU(N) gauge theory is not consistent. These problems have been overcome by Wess {\it et. al.} \cite{Jurco} who constructed nonabelian
gauge theory on the enveloping Lie algebra resulting from the Moyal star product. No extra fields other than fields in ordinary commutative gauge field theory 
appear in their formulation after performing the Seiberg-Witten map \cite{SW}.
This approach has allowed them to construct the noncommutative standard model
as well as SO(10) GUT \cite{Jurco2}. 
NC standard model is also constructed by \cite{CPST} in the different approach.
\par
In this article, 
we will
formulate the NC nonabelian gauge theory by use of the enveloping
gauge group SU(N)$\ast$ with enveloping Lie algebra 
which reduces to ordinary SU(N) gauge group in the limit of 
NC parameter $\theta^{\mu\nu}\to0$. Seiberg-Witten map 
\cite{SW} doesn't play an important role as in Wess 
{\it et.al.}\cite{Jurco}.
Thus, our NC non abelian gauge theory is different from
that of  Wess {\it et. al.} 
\cite{Jurco}. 
Our nonabelian gauge theory has no $A^0_\mu(x)$ component since it is only induced 
by gauge transformation and can be eliminated away. Thus, we don't need 
to handle additional fields.
\par
After the construction of standard model in particle physics more than 
decades ago, there have been several occasions  that indicated
the experimental deviations from the standard model. However, 
such deviations ultimately shrank to nothing and the correctness of 
the standard model has been confirmed. Thus, though there are many advanced 
theories beyond standard model, they must reduce to the standard model in their characteristic limits. This is the case also 
in the NC field theory.
If not, it is branded not to be the qualified theory beyond standard model.
The indication by Hayakawa \cite{Hayakawa} is serious to the
NC gauge theory since 
quarks in standard model have fractional charges. There must be 
the story other than Hayakawa's indication if the noncommutativity
on the spacetime is somewhat true in nature. 
We overcome this difficulty by
considering nonabelian SU(N)$\ast$ gauge theory in which the 0 component
$A^0_\mu(x,\theta)$ of gauge field is induced by gauge transformation, and 
the spontaneous symmetry breakdown of such gauge theory.
 Then, referring to this case, we propose the NC 
gauge theory with arbitrary U(1) charge.
We conclude that such gauge theory 
can not exist alone, but it must coexist with 
NC nonabelian gauge theory.\par
This article consists of 5 sections. In second section, the
NC nonabelian gauge theory is proposed. 
In third section, the spontaneous symmetry breakdown of SU(4)$\ast$ gauge
theory is discussed in order to show that quarks has the fractional 
$B$-charges whereas the lepton has charge $-1$. 
In fourth section, the gauge theory with arbitrary charge is
proposed referring to the example in third section. The last section
is devoted to short conclusions.

\section{Nonabelian gauge theory on noncommutative spacetime\label{S2}}
Let us first consider the nonabelian gauge theory 
on the NC spacetime
with the symmetry SU(N)  given by the Lagrangian
\begin{align}
{\cal L}=&-\frac12\text{Tr}\left[F_{\mu\nu}(x)\ast F^{\mu\nu}(x)\right]
+{\mit\bar\Psi}(x)\ast\{i\gamma^\mu(\partial_\mu-igA_\mu(x))-m\}
\ast{\mit\Psi}(x)\nonumber\\
&+\text{Tr}\left[({\cal D}^\mu\varphi(x))^\dag\ast {\cal D}_\mu\varphi(x)
+m^2\varphi(x)^\dag \ast\varphi(x)-\lambda (\varphi(x)^\dag \ast\varphi(x))^2
\right],
\label{2.1}
\end{align}
where we omit the gauge fixing and FP ghost terms. 
The Moyal star product of functions $f(x)$ and $g(x)$ is 
defined as
\begin{equation}
f(x)\ast g(x)=\left.e^{\frac{i}{2}\theta^{\mu\nu}\partial^1_\mu\partial^2_\nu}f(x_1)g(x_2)\right|_{x_1=x_2=x},
\end{equation}
where $\theta^{\mu\nu}$ is two rank tensor 
to characterize the noncommutativity of
spacetime.  Though  $\theta^{\mu\nu}$ is usually seemed to be 
a constant not to transform corresponding to Lorentz transformation,
$\theta^{\mu\nu}$ is regarded as  two rank 
tensor in this letter as discussed in \cite{IMO}.
$\mit\Psi(x)$ is the fermion field with the fundamental representation.
The quantity
\begin{equation}
F_{\mu\nu}(x)=\partial_\mu A_\nu(x)-\partial_\nu A_\mu(x)-ig\,[A_\mu(x),\,A_\nu(x)]_*\label{2.2}
\end{equation}
is the field strength of gauge field 
with the configuration
\begin{equation}
A_\mu(x)=\sum_{a=0}^{N^2-1}A^a_\mu(x)T^a.\label{2.3}
\end{equation}
and the field  $\varphi(x)$ is the Higgs boson belonging to the adjoint 
representation of SU(4) which covariant derivative is  
\begin{equation}
{\cal D}^\mu\varphi(x)=\partial^\mu\varphi(x)-ig[A^\mu(x),\,\varphi(x)]_\ast.
\end{equation}
\par
\par
 The gauge transformations of fields in \eqref{2.1} are defined as
\begin{align}
\begin{aligned}
& A^g_\mu(x)=U(x,\theta)\ast A_\mu(x)\ast U^{-1}(x,\theta)+\frac{i}{g}U(x,\theta)\ast\partial_\mu U^{-1}(x,\theta),\\
&{\mit\Psi}^g(x)=U(x,\theta)\ast{\mit\Psi}(x),\\
&\varphi^{\,g}(x)=U(x,\theta)\ast \varphi(x)\ast U^{-1}(x,\theta)
\label{2.6}
\end{aligned}
\end{align}
where the gauge transformation function $U(x,\theta)$ is written as
\begin{equation}
U(x,\theta)=e^{i\alpha(x,\theta)\ast}=\sum_{n=0}^\infty \frac{i^n}{n!}\;\alpha(x,\theta)\ast\alpha(x,\theta)\ast\alpha(x,\theta)\ast\cdots\ast\alpha(x,\theta)
\label{2.9}
\end{equation}
in terms of the Lie algebra valued function
\begin{equation}
\alpha(x,\theta)=\sum_{a=0}^{N^2-1}\alpha^a(x,\theta)\,T^a\label{2.10}
\end{equation}
with the condition that
\begin{equation}
\lim_{\theta\to0}\alpha^0(x,\theta)=0.
\label{2.11}
\end{equation}
We call the ensemble of Lie algebra valued functions \eqref{2.10} 
keeping the condition \eqref{2.11} enveloping Lie algebra.
In similar way, the ensemble of gauge function \eqref{2.9} is
enveloping gauge group denoted by SU(N)$\ast$.
The star commutator between two Lie algebra valued functions 
is calculated as
\begin{align}
[\alpha(x,\theta),\,\beta(x,\theta)]_{\ast}=&\sum_{a,\,b=0}^{N^2-1}
\left(\alpha^a(x,\theta)\ast\beta^b(x,\theta)T^aT^b-\beta^b(x,\theta)\ast\alpha^a(x,\theta)T^b
T^a\right)
\nonumber\\
=&\sum_{c=0}^{N^2-1}\sum_{a,b=0}^{N^2-1}\left(if^{abc}\,\frac12\{\alpha^a(x,\theta),\,\beta^b(x,\theta)\}_{\ast}+d^{abc}\,\frac12[\alpha^a(x,\theta),\,\beta^b(x,\theta)]_{\ast}\right)T^c,
\label{2.12a}
\end{align}
where
\begin{equation}
[T^a,\,T^b]=\sum_{c=0}^{N^2-1}if^{abc}T^c,\hskip5mm
\{T^a,\,T^b\}=\sum_{c=0}^{N^2-1}d^{abc}T^c,\hskip5mm
\text{Tr}\left(T^aT^b\right)=\frac{1}{2}\delta^{ab}.
\end{equation}
Thus, owing to the condition \eqref{2.11}, the enveloping
Lie algebra closes within itself for the star commutator \eqref{2.12a}.
It should be noted that 
\begin{equation}
\lim_{\theta\to0}U(x,\theta)=U(x,\theta)=e^{i\sum_{a=1}^N\alpha^a(x)T^a}
\in \text{SU(N)},\label{2.14}
\end{equation}
which indicates that the enveloping group SU(N)$\ast$ reduces to nonabelian
group SU(N) when $\theta^{\mu\nu}$ becomes to 0.
\par
Under the gauge transformation in \eqref{2.6} 
the field strength $F_{\mu\nu}(x)$ and
the covariant derivative of ${\cal D}_\mu\varphi(x)$
are transformed covariantly
\begin{align}
&F_{\mu\nu}^g(x)=U(x,\theta)\ast F_{\mu\nu}(x)\ast U^{-1}(x,\theta),\\
&({\cal D}_\mu\varphi(x))^g=U(x,\theta)\ast 
{\cal D}_\mu\varphi(x)\ast U^{-1}(x,\theta).\label{2.16}
\end{align}
Then, the gauge field term in \eqref{2.1} is transformed as in
\begin{equation}
\text{Tr}\left[F_{\mu\nu}^g(x)\ast {F^g}^{\mu\nu}(x)\right]
=\text{Tr}\left[U(x,\theta)\ast F_{\mu\nu}(x)\ast F^{\mu\nu}(x)
\ast U^{-1}(x,\theta)\right]
\label{2.22a}
\end{equation}
which shows the gauge term itself is not gauge invariant because of the 
Moyal $\ast$product but the action is invariant thanks to the rule
\begin{equation}
\int d^4x \;f(x)\ast g(x)=\int d^4x \;g(x)\ast f(x).
\end{equation}
We call this situation 
 pre-invariance to gauge transformation. That is, 
the gauge boson term in Lagrangian is pre-invariant. 
That is the case for the Higgs boson term in \eqref{2.1} because 
 of \eqref{2.16}. 
On the other hand, the fermion term in Eq.\eqref{2.1} is invariant under gauge transformations \eqref{2.6}. 
\par
In order to construct the nonabelian gauge theory on the NC 
spacetime, we start from \eqref{2.1} where the 0 component $A_\mu^0(x,\theta)$
 of the gauge field $A_\mu(x)$ doesn't get in.
But, it is induced by the gauge transformation \eqref{2.6}.
Thus, the 0 component of ${A^g}_\mu(x)$ in \eqref{2.6}
depends on  other components $A^a_\mu(x)\;(a=1,2,\cdots,15)$ 
and so, it is not independent field. 
Moreover it vanishes  owing to  \eqref{2.10} 
when the NC parameter $\theta^{\mu\nu}$
approaches to 0. Thus, 
we doesn't need to quantize $A_\mu^0(x,\theta)$.
Thus, even if the 0 component of  ${A^g}_\mu(x)$ appears
in \eqref{2.6}, it is out of our considerations because we
eliminate it away by the gauge transformation.
The existence of the enveloping SU(N)$\ast$ gauge group
insures the construction of nonabelian gauge theory on the NC 
spacetime.
\par
Let us explain the difference between the nonabelian NC gauge theory in this
paper and that of Wess et.al. \cite{Jurco}.
According to the paper of Wess et.al. \cite{Jurco}
they construct the nonabelian gauge theory based on 
the enveloping Lie algebra restricted to satisfy
\begin{equation}
(\delta_\alpha\delta_\beta-\delta_\beta\delta_\alpha)=
\delta_{\alpha\times\beta},\label{A}
\end{equation}
where
\begin{equation}
\begin{aligned}
&\delta_\alpha\psi(x)=i\Lambda(x)\ast\psi(x)\\
&\delta_\alpha A_\mu(x)
=\partial_\mu \Lambda(x)+i[\Lambda(x),\; A_\mu(x)]_{\ast}
\end{aligned}
\end{equation}
with $\Lambda(x)$ being an element of the enveloping Lie algebra.
In order to satisfy \eqref{A}, $\Lambda(x)$ must be the function of 
ordinary gauge field $a_\mu(x)$ and the Lie algebra $\alpha(x)$
satisfying
\begin{equation}
i\delta_\alpha\Lambda_\beta[a]-i\delta_\beta\Lambda_\alpha[a]
+\Lambda_\alpha[a]\ast \Lambda_\beta[a]-\Lambda_\beta[a]\ast \Lambda_\alpha[a]
=i\Lambda_{\alpha\times\beta}[a]
.\label{C}
\end{equation}
The gauge field $A_\mu(x)$ and fermion field $\psi(x)$ must depend on 
$a_\mu(x)$.
Then, they expand these function and fields in the order of $h\,\theta^{\mu\nu}$.
They call this process Seiberg-Witten map.
According to their expansions,
\begin{equation}
\begin{aligned}
&\Lambda_\alpha[a]=\alpha(x)+h\Lambda^1_\alpha[a]
+h^2\Lambda^2_\alpha[a]+\cdots,\\
&\psi[a]=\psi^0[a]+h\psi^1[a]+h^2\psi^2[a]+\cdots,\\
&A_\mu[a]=a_\mu(x)+hA^1_\mu[a]+h^2A^2_\mu[a]+\cdots,
\end{aligned}
\end{equation}
where
terms higher than the order $h^1$ are the very complicated functions of
$\alpha(x),\;a_\mu(x)$ and their derivatives.
Action which consists of gauge and fermion fields is also expanded in 
the series of $h\,\theta^{\mu\nu}$ as in Eqs.(5.4) and (5.5) in \cite{Jurco}.
The 0-th order in the expansion is the ordinary action composed of ordinary fields $a_\mu(x),\;\psi^0(x)$.
Each term in expansions is invariant under ordinary gauge transformation.
Terms higher than $h^1$ in the expansion of action are very complicated with 
$F_{\mu\nu}^0,\; \psi^0$ and their derivatives. Their theory looks 
unrenormalizable because the couplings have the higher dimensions. 
\par
The gauge transformation function $U(x,\theta)$ in this paper is written 
\eqref{2.9}
in terms of the Lie algebra valued function $\alpha(x,\theta)$
with the condition \eqref{2.11}.
In this paper,
we do not claim the restriction \eqref{A}, and therefore, 
we do not need the Seiberg-Witten map. 
Gauge field $A_\mu(x)$ and fermion field $\psi(x)$ as well as the action are not expanded in $\theta^{\mu\nu}$. 
Thus, it is concluded that our non-abelian gauge field on NC
spacetime is much different from that of Wess et.al. 
\cite{Jurco}.

\section{The spontaneous breakdown of SU(4)$\ast$ gauge theory}
SO(10) grand unified theory (GUT) including its supersymmetric version 
is most promising model in particle physics 
since it can incorporate the 15 existing fermions in addition to 
the right-handed neutrino and has possibilities to explain so many
phenomenological puzzles. Pati-Salam symmetry $\text{SU(4)}\times \text{SU(2)}_L\times \text{SU(2)}_R$ is one of the intermediate symmetry of the spontaneous breakdown of SO(10) GUT. This symmetry spontaneously breaks down to
the left-right symmetric gauge model with the symmetry 
$\text{SU(3)}_c\times \text{SU(2)}_L\times \text{SU(2)}_R
\times \text{U(1)}_B$.
In this stage, the spontaneous breakdown 
$\text{SU(4)}\to \text{SU(3)}_c\times \text{U(1)}_B$
occurs. $B$ charge of fermions is given by
\begin{equation}
Q_B
\begin{pmatrix}
q^r \\ q^g \\ q^b \\ l
\end{pmatrix}
=
\begin{pmatrix}
\frac13 \\[0.3mm] \frac13 \\[0.3mm] \frac13 \\[0.3mm] \text{\small$-{1}$}
\end{pmatrix}
\begin{pmatrix}
q^r \\ q^g \\ q^b \\ l
\end{pmatrix}.\label{3.1}
\end{equation}
As an example, we pick up this process in order to investigate
whether fermions have $B$ charge in \eqref{3.1} in the noncommutative 
version of SU(4) gauge theory.
\par
The gauge boson in SU(4)$\ast$ gauge theory is expressed 
in terms of 16 component gauge bosons by
\begin{align}
A_\mu(x)&=\sum_{a=0}^{15}A_\mu^a(x)\,T^{\,a}\nonumber\\[2mm]
&=\frac{1}{{2}}\begin{pmatrix}
A_\mu^{11}& G^{1}_\mu & G^{2}_\mu & X^{1}_\mu \\[2mm]
{\bar G}^{1}_\mu & A_\mu^{22}
 & G^{3}_\mu & X^{2}_\mu \\[2mm]
{\bar G}^{2}_\mu & {\bar G}^{3}_\mu & A_\mu^{33} & X^{3}_\mu \\[2mm]
{\bar X}^{1}_\mu & {\bar X}^{2}_\mu & {\bar X}^{3}_\mu & 
A_\mu^{44}\\
\end{pmatrix}.
\end{align}
where
\begin{align}
&\left\{\begin{aligned}
&A_\mu^{11}=\dfrac{1}{\sqrt{2}}A^0_\mu+A^{3}_\mu
+\dfrac{1}{\sqrt{3}}A^{8}_\mu+\dfrac{1}{\sqrt{6}}A^{15}_\mu ,\\
&A_\mu^{22}=\dfrac{1}{\sqrt{2}}A^0_\mu
-A^{3}_\mu+\dfrac{1}{\sqrt{3}}A^{8}_\mu+\dfrac{1}{\sqrt{6}}A^{15}_\mu\\
&A_\mu^{33}=\dfrac{1}{\sqrt{2}}A^0_\mu
-\dfrac{2}{\sqrt{3}}A^{8}_\mu
+\dfrac{1}{\sqrt{6}}A^{15}_\mu\\
&A_\mu^{44}=\dfrac{1}{\sqrt{2}}A^0_\mu-\dfrac{3}{\sqrt{6}}A^{15}_\mu
\end{aligned}\right.\\
&G_\mu^1=A_\mu^1-iA_\mu^2,\hskip5mm G_\mu^2=A_\mu^4-iA_\mu^5,\hskip5mm
G_\mu^3=A_\mu^6-iA_\mu^7,\\
&X_\mu^1=A_\mu^9-iA_\mu^{10},\hskip5mm X_\mu^2=A_\mu^{11}-iA_\mu^{12},\hskip5mm
X_\mu^3=A_\mu^{13}-iA_\mu^{14}.
\end{align}
The gauge field $A_\mu(x)$ 
contains 8 color gluons, 6 gauge bosons causing proton decay,  
one extra boson $A^{15}_\mu(x)$, and one 0 component boson $A^{0}_\mu(x)$ dependent on other bosons. Here, we denote $A^{15}_\mu(x)$ by $B_\mu(x)$ and call it 
$B$-field.

The vacuum expectation value of Higgs boson $\varphi(x)$ takes the form
\begin{equation}
<\varphi(x)>=v
\begin{pmatrix}
1 & 0 & 0 & 0 \\
0 & 1 & 0 & 0 \\
0 & 0 & 1 & 0 \\
0 & 0 & 0 & -3 
\end{pmatrix}
\label{3.3}
\end{equation}
which yields the gauge boson mass term
\begin{align}
\left|-ig[A_\mu(x),\,<\varphi(x)>]\rule{0mm}{3.3mm}\right|^2&=
{8g^2v^2}\left({X^1}\rule{0mm}{2mm}^\mu {\bar X}_\mu^1
+{X^2}\rule{0mm}{2mm}^\mu {\bar X}_\mu^2
+{X^3}\rule{0mm}{2mm}^\mu {\bar X}_\mu^3\right).\label{3.4}
\end{align}
Equation \eqref{3.4} shows that 6 proton decay causing gauge bosons acquire
mass, so that symmetries relating to Lie algebra $T^a\;(a=1,2,\cdots,8,15)$
keep unbroken. Let us consider the gauge transformation specified by
\begin{align}
U^{cb}(x,\theta)=e^{i\alpha(x,\theta)\ast}
=\exp i\left\{{\sum_{a=0}^8T^a\alpha^a(x,\theta)\ast+
T^{15}\alpha^{15}(x,\theta)\ast}\right\}.
\label{3.5}
\end{align}
Under this gauge transformation,
the part of the gauge field $A_\mu(x)$ in \eqref{2.3}
\begin{align}
A_\mu^{cb}(x)={\sum_{a=0}^8A^a_\mu(x)\,T^a+
B_\mu(x)T^{15}}
\end{align}
and fermion field $\mit\Psi(x)$ and Higgs field transform in the similar way as in \eqref{2.6}.
Thus, it is easily shown that
the Lagrangian \eqref{2.1} is still pre-invariant after the spontaneous
breakdown coming from \eqref{3.3}.
This indicates that color symmetry yielding the strong interaction and 
$B$-symmetry due to the generator $T^{15}$ remain unbroken.
In the commutative field theory, this breakdown is written as
\begin{align}
\text{SU(4)} \rightarrow \text{SU(3)}\times \text{U(1)}.
\end{align}
However, we can't do it in the same way because 
\begin{align}
U^{cb}(x,\theta)\ne 
\exp i\left\{\sum_{a=1}^8\alpha^a(x,\theta)\ast\,T^a\right\}
\ast
\exp i\left\{{
\alpha^{15}(x,\theta)\ast\,T^{15}}\right\}
\label{3.8}
\end{align}
owing to the Moyal product. Thus, in the NC field theory,
we should write the spontaneous breakdown explained so far as
\begin{align}
\text{SU(4)}\ast \rightarrow \text{SU(3$+$1)}\ast.
\end{align}

\par
Interaction terms between fermion and $B$-gauge field extracted from the fermion term in \eqref{2.1} is given by
\begin{align}
{\cal I}_D&={\mit\bar\Psi}(x)\ast\{\gamma^\mu(gB_\mu(x)T^{15})\}
\ast{\mit\Psi}(x)\nonumber\\
&=\frac{3}{2\sqrt{6}}\,g\,{\mit\bar\Psi}(x)\ast\gamma^\mu
\begin{pmatrix}
\frac13 & 0 & 0 & 0 \\
0 & \frac13 & 0 & 0 \\
0 & 0 & \frac13 & 0 \\
0 & 0 & 0 & \text{\small$-1$} 
\end{pmatrix}
B_\mu(x)
\ast{\mit\Psi}(x)
\end{align}
Then, if we define $B$-charge operator $Q_B$ 
and $B$-charge $e_{\text{\tiny$B$}}$
\begin{align}
Q_B=\frac{2\sqrt{6}}{3}T^{15},\hskip1cm 
e_{\text{\tiny$B$}}=\frac{3}{2\sqrt{6}}\,g
\end{align}
\eqref{3.1} is reproduced.
\section{Arbitrary $B$-charge}
In the previous section, we considered the spontaneous breakdown
of SU(4)$\ast$ gauge symmetry down to SU(3+1)$\ast$ symmetry. 
Thus, charges of fermions are limited as shown in \eqref{3.1}.
However, apart from such construction, we can considered  such a case that 
if the Lagrangian is pre-invariant under the gauge transformation function $U^s(x,\theta)$ given by
\begin{align}
U^{s}(x,\theta)=e^{i\alpha(x,\theta)\ast}
=\exp i\left\{\sum_{a=0}^8\,T^a\,\alpha^a(x,\theta)\ast+
Q\,\beta(x,\theta)\ast\right\}
\label{4.1}
\end{align}
where
\begin{align}
Q=
\begin{pmatrix}
e & 0 & 0 & 0 \\
0 & e & 0 & 0 \\
0 & 0 & e & 0 \\
0 & 0 & 0 & e' 
\end{pmatrix}
\end{align}
with arbitrary constants $e$ and $e'$, fermions may have arbitrary 
charges. This is because 
Interaction terms between fermion and $B$-gauge is given by
\begin{align}
{\cal I}_D&={\mit\bar\Psi}(x)\ast\{\gamma^\mu(gB_\mu(x)Q)\}
\ast{\mit\Psi}(x).
\end{align}
\par
If there is only U(1)$\ast$ gauge symmetry, the gauge transformation 
of gauge field 
$A_\mu(x)=QB_\mu(x)$
given by
\begin{align}
U^{e}(x,\theta)
=\exp i\left\{
Q\,\beta(x,\theta)\ast\right\}.
\label{4.3}
\end{align}
leads to the inconsistency as indicated by Hayakawa \cite{Hayakawa}. 
But, in our case, there are two kinds of symmetry and therefore, 
the gauge transformation of gauge field 
$A_\mu(x)=\sum_{a=0}^8T^a{A^a}_\mu(x,\theta)+
QB_\mu(x)$ given by \eqref{4.1} has nothing to do with any contradiction because of $A_\mu^0(x,\theta)$ existence.
\section{Conclusions}
We have proposed nonabelian SU(N)$\ast$ gauge theory on NC 
spacetime which doesn't depend on
the Seiberg-Witten map \cite{SW} as in Jurco {\it et.al.} \cite{Jurco}. 
We considered  the SU(4)$\ast$ gauge theory 
which spontaneously breaks down to SU(3$+$1)$\ast$ symmetry
in order to obtain the gauge theory with plural U(1) charges. 
It is shown that such NC gauge theory 
with different charges more than two  
can not exist alone, but it must coexist with 
NC nonabelian gauge theory.
\par
Let us discuss the present situations on NC gauge theory and show 
several undesirable results with respect to the quantized version 
of NC gauge theory. 
Mart$\grave{\text{i}}$n and  S$\acute{\text{a}}$nchez-Ruiz
 \cite{M-SR}  showed that
U(1) NC gauge theory can be renormalizable but asymptotically free. 
Matusis, Susskind and Toumbas \cite{MST} found the unfamiliar IR/UV connection
in NC gauge theory which involves non-analytic behavior in NC parameter
$\theta$ making the limit $\theta\to0$ singular.
It is also pointed out by Armoni \cite{Armoni}
that the $\theta\to0$ limit of the U(N) NC gauge theory
does not converge to the ordinary SU(N)$\times$U(1) gauge theory.
Finally, Ruiz Ruiz \cite{Ruiz} indicated tachyonic instability in
U(1) NC gauge theory due to UV/IR mixing effects.
In this paper, we do not investigate the quantum effects of the SU(N)$\ast$
gauge theory. Then, it may display the same undesirable points which should 
be quickly addressed. However, 
we construct in this paper  new formulation of non-abelian gauge theory 
on NC spacetime and propose how to incorporate 
fields with the arbitrary U(1) charges 
into NC gauge theory. These are main purposes in this paper which
 is the first step in order to achieve the true theory. 
As showed by Ruiz Ruiz \cite{Ruiz} and authors of \cite{MST},
the defects stated above may be solved 
by considering the supersymmetric version
of NC gauge theory. Then, we must construct the supersymmetric version 
of the present paper. However, there is a possibility to solve 
the problems within the present formalism
since  it has the different aspects from other NC gauge theories stated above.
In this paper, $A_\mu^0$ component field  is the dependent field on
other components gauge fields and it can be eliminated away
 by gauge transformation.
The charges in this paper are related to the $\lambda^{15}$ in SU(4)$\ast$
 NC gauge theory with spontaneous breakdown which does not need to take 
the limit of $\theta\to0$.  These possibilities will be investigated in future
works.
\par
The deviation from the standard model in particle physics has not yet observed,
and so any model beyond standard model must reduce to it in 
some approximation. NC gauge theory must also reproduce standard
model in the limit of NC parameter $\theta^{\mu\nu}\to0$. 
According to the considerations in section 3, 
the eligible
enveloping gauge group to construct the standard model on NC spacetime seems to be SU($3_c+2_L+1_Y$)$\ast$. This work will appear in our forthcoming paper.


\end{document}